\documentclass[a4paper,11pt]{article}

\usepackage{graphicx} 
\usepackage{longtable,array}
\usepackage{rsc}
\usepackage{epsf}
\usepackage{color}
\usepackage{soul}

\newcommand{\etal}{{\it et al.} }
\newcommand{\ai}{{\it ab initio}}

\newcommand{\cm}{cm$^{-1}$}

\newcommand{\hp}{H$_3^+$}
\newcommand{\ho}{H$_2$O$_2$}

\newcommand{\onlinecite}[1]{\hspace{-1 ex} \nocite{#1}\citenum{#1}}

\begin{document}

\title{Variational calculation of highly excited rovibrational energy
levels of  \ho }

\author
{Oleg L. Polyansky\thanks{
Department of Physics and Astronomy, University College London,
Gower Street, London WC1E 6BT, United Kingdom}
\thanks{Institute of Applied Physics, Russian Academy of Science,
Ulyanov Street 46, Nizhny Novgorod, Russia 603950
E-mail: \texttt{o.polyansky@ucl.ac.uk}},
Igor N. Kozin$^\dagger$, Roman I. Ovsyannikov$^\dagger$,
\and
Pawe\l\  Ma{\l}yszek$^\ddagger$, Jacek Koput\thanks{Department of Chemistry, Adam Mickiewicz University, Poznan, Poland},
Jonathan Tennyson$^{\ast}$ and Sergei N. Yurchenko$^{\ast}$
}

\noindent\textit{\small{\textbf{Received Xth XXXXXXXXXX 20XX, Accepted Xth XXXXXXXXX 20XX\newline
First published on the web Xth XXXXXXXXXX 200X}}}

\noindent \textbf{\small{DOI: 10.1039/b000000x}}
\vspace{0.6cm}


\maketitle

\begin{abstract}
  Results are presented for highly accurate \ai\ variational calculation of the
  rotation - vibration energy levels of \ho\ in its electronic ground
  state. These results use a recently computed potential
  energy surface and the  variational nuclear-motion programmes WARV4,
  which uses an exact kinetic energy (EKE) operator, and TROVE, which
  uses a                     numerical expansion for the kinetic energy.  The
  TROVE calculations are performed for levels with high values of
  rotational excitation, $J$ up to 35.  The purely \ai\ calculations
  of the rovibrational energy levels reproduce the observed
  levels with a standard deviation of about 1 \cm, similar to that of
  the $J = 0$ calculation as the discrepancy between
  theory and experiment for rotational energies within a given
  vibrational state is substantially determined by the error in the
  vibrational band origin. Minor adjustments are made to
  the \ai\ equilibrium geometry and to the height of the torsional
  barrier.  Using these and correcting the band origins using the
  error in $J = 0$ states           lowers the standard
  deviation of the observed $-$ calculated energies
 to      only  0.002 \cm\ for levels up to $J = 10$ and 0.02 \cm\
  for all  experimentally know energy levels, which extend up to $J = 35$.
\end{abstract}

\section{Introduction}
\label{s:intro}

Hydrogen peroxide (\ho) is a well-studied system because of its
unusual properties, particularly the almost freely rotating
OH moieties.
The role of \ho\ in the chemistry of the Earth's atmosphere \cite{74Davis.H2O2,03ViMaMi.H2O2, 13AlAbBeBo.H2O2} as well as of the Martian atmosphere \cite{04EnBeGr.HOOH,12EnGrLe.HOOH}  has been widely acknowledged. It has also
recently been detected in the interstellar medium \cite{11BePaLi.H2O2}.
The vibration-rotation spectra of hydrogen peroxide has attracted
significant attention, both experimental
and theoretical. For example it has been used
                                                    as a benchmark system with large amplitude motion for
testing different variational nuclear motion codes \cite{93.BrCarrin.HOOH,00CaHaxx.HOOH,00Luckhaus.HOOOH,01ChMaGu.HOOH,02YuMuck.HOOH,02Mladenovic.HOOH,03LiGuxx.HOOH,09CaHaBo.h2o2,11CaShBo.H2O2}.


Very recently Ma{\l}yszek and Koput \cite{Koput3} presented a highly accurate \ai\
potential energy surface (PES) of HOOH which was shown to reproduce the known vibrational band origins
with the average accuracy of 1 \cm. Other \ai\ PESs of HOOH were reported by Harding \cite{91Harding.HOOH},
Kuhn \etal\cite{99Kuhn.HOOH}, Senent \etal \cite{00Senent.HOOH}, and Koput \etal\cite{Koput1}.

Hydrogen peroxide has three important properties from the viewpoint of
variational calculations. Firstly, the large amplitude motion
of the OH internal rotors that has already been mentioned. Secondly, its
relatively low dissociation energy of about 17,000 \cm\
has made \ho\ a benchmark tetratomic molecule for experimental study
of the dissociation process \cite{Rizo}. Thirdly, \ho\  is a tetratomic
system where variational calculations can really aid the analysis
of spectra.

For triatomic molecules, accurate
calculation of the rotation-vibration levels to high
accuracy using variational nuclear motion
methods has become routine \cite{jt309,jt512,12HuScTaLe.CO2}. For tetratomic
molecules this process is just beginning; it is natural for initial
high accuracy studies to focus on molecules with large
amplitude motion such as  ammonia \cite{Huang:Schwenke:Lee:2010:I,Huang:Schwenke:Lee:2010:II} and hydrogen
peroxide.

The advantages of using variational calculations to assign
vibration-rotation spectra of triatomic molecules has been
demonstrated for several molecules. Initial studies focused
on \hp\ \cite{jt80,jt102,jt193} and water
\cite{jt200,jt218}, systems for which the use of variational
calculations to analyse spectra is now the accepted procedure.
In particular, spectra involving hot molecules, and hence high rotational
states, and large amplitude motion,
such as \hp\ on Jupiter \cite{jt80}  and water on the Sun
\cite{jt200}, assignments using the traditional, effective
Hamiltonian approach are almost impossible.

A significant advantage of variational calculations over effective
Hamiltonian techniques is the automatic allowance for accidental
resonances between vibrations. Whereas for most triatomic molecules such
resonances become significant at fairly high vibrational
energies, for tetratomic molecules accidental resonances
can even make the analysis of low-lying vibrational states intractable using effective Hamiltonians. \ho\ is a good example of this situation.  Although \ho\ spectral lines
are strong and were first observed more than seventy years ago
with spectrometers  much less sophisticated then those available
nowadays \cite{41ZuGi.H2O2,50Giguer.H2O2}, the analysis of experimental spectra involving high $J$ transitions for \ho\ is only complete up to
2000 \cm\ \cite{perin1,perin2,perin3}, significantly lower
in frequency than transitions to the OH stretching fundamentals.
One reason for this is the
complication of the analysis by accidental resonances.
Accurate variational calculations on \ho\ offer a way out of this impasse.


Recent advances in variational calculations suggest that they can be
used for  systems larger than triatomic.  High accuracy variational
calculations of the spectra and line lists for tetratomic molecules
such as ammonia \cite{jt466,jt500,Huang:Schwenke:Lee:2010:II} have been
performed.
         These NH$_3$ line lists have been used both for the assignment
of transitions involving higher vibrational states \cite{12SuBrHu.NH3},  hot rovibrational spectra involving high $J$ levels \cite{jt508}
as well as  for correcting and improving the analysis
of more standard transitions \cite{Huang:Schwenke:Lee:2010:II,jt543}.  Numerical
calculations of wavefunctions for high $J$ states of tetratomic molecules
are possible not only because modern computers have the ability to diagonalise
larger matrices but also because, as illustrated below, the accuracy
of calculations employing
approximate kinetic energy operators \cite{multimode,trove-paper}
becomes comparable with those using an exact kinetic energy approach \cite{jt312}.  While high $J$ calculations within the exact kinetic
energy approach  are still
computationally challenging for tetratomic molecules,
calculations with $J \sim 50$ are feasible with approaches such as
TROVE \cite{trove-paper}.  Furthermore, the possibility of calculating \ai\
dipole moment surfaces of extremely high accuracy \cite{jt509}
enhances the value of using variational calculations since they can also be
used to
create line lists.
These factors raise the possibility of creating accurate line lists for
\ho.  However, the presence of the large amplitude, torsional motion
of the two OH fragments in \ho\ complicates the problem. This requires an appropriate
nuclear motion programme for calculation of the
rovibrational energy levels by solving the corresponding Schr\"odinger
equation; this programme should be able to compute high $J$ levels
within the limitations of the modern computers.

In this paper we compute high accuracy rovibrational energy
levels going to high $J$ for
\ho\  using the                          \ai\       PES due to
Ma{\l}yszek and Koput \cite{Koput3}.
 To do this we
test two nuclear motion programmes:
the exact kinetic energy (EKE)
programme WAVR4 \cite{jt339}  and approximate kinetic energy programme TROVE
\cite{trove-paper}. It is  shown that use of
TROVE allows us to calculate very high $J$ energy levels which are
in excellent agreement with observation.
 The paper is organised as follows.
Section~\ref{s:method}  describes the modifications of the TROVE programme
necessary to make it suitable for the calculation of spectra
of such a nonrigid molecule.
Section~\ref{s:results} describes the details of computations performed.
Section~\ref{s:concl} presents our results which is followed by the concluding section
which discusses prospects for further work on this system.

\section{Methods of calculation }
\label{s:method}

\begin{figure}
\begin{center}
 \leavevmode
\epsfxsize=7.0cm \epsfbox{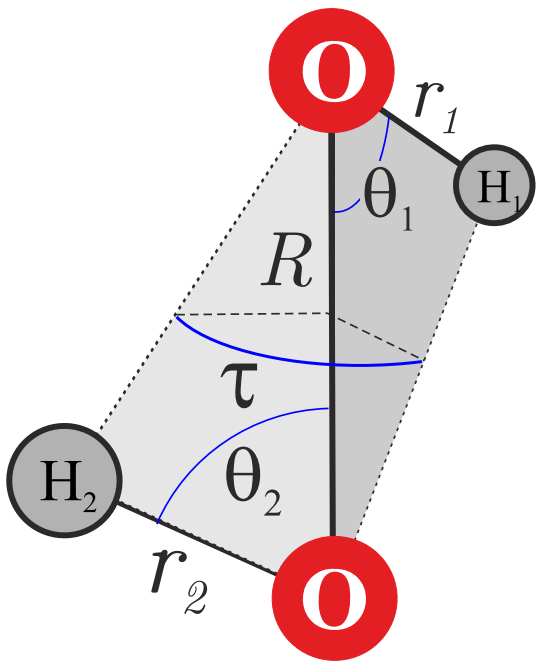}
\vspace*{0.5cm}
\caption{\label{f:opt:geom} The internal coordinates for the HOOH molecule. }
\label{f:tau}
\end{center}
\end{figure}

The accuracy of a calculation of  rovibrational energy levels depends
first of all on the accuracy of the potential energy surface (PES) used
as input to the nuclear motion Schr\"odinger equation. Until recently the most
accurate PES for \ho\ was  the one due to Koput \etal\cite{Koput1} which gave a
typical discrepancy between theory and experiment for vibrational band
origins of about 10 \cm\ \cite{Koput2}.  However,
two of us\cite{Koput3} recently determined a very accurate PES computed using the explicitly correlated coupled-cluster method  [CCSD(T)-F12] method,
\cite{CC,CC2} in the F12b form \cite{F12a} as implemented in
the  MOLPRO package \cite{molpro}.
Various correlation-consistent basis sets were used for various parts of the PES, the largest being aug-cc-pV7Z. The  CCSD(T)-F12 results were augmented with the Born-Oppenheimer diagonal, higher-order valence-electron correlation, relativistic, and core-electron correlation corrections. The 1762 \ai\ points obtained were fitted to the functional form
\begin{equation}
\label{e:koput}
V(q_1,q_2,q_3,q_4,q_5,q_6) = \sum_{ijklmn} c_{ijklmn} q_1^{i} q_2^{j}q_3^{k} q_4^{l} q_5^{m} \cos{n q_6}
\end{equation}
where $q_{i}$  ($i = 1,2,3$)  are the Simons-Parr-Finlan stretching OO and OH coordinates  \cite{73SimParrFin.SPF} $q_1= (R-R_{\rm e})/R $ and
$q_i = (r_i - r_{\rm e})/r $ ($i=1,2$), $q_4 = \theta_1 - \theta_{\rm e} $ and $q_5 = \theta_2 - \theta_{\rm e} $ are the two OOH bending coordinates,  $q_6 = \tau$ is the torsional angle $\angle$HOOH (see Fig.~\ref{f:tau}),
and $R_{\rm e}$, $r_{\rm e}$, and $\theta_{\rm e}$ are the
corresponding equilibrium values. The
expansion coefficients  $c_{ijklmn}$  used in this work are
given in the supplementary material \cite{supl} to this
article (see also Ref. \citenum{Koput3}).

Recent calculations\cite{Koput3} using this PES gave, for the 30 observed vibrational band origins of \ho, a standard deviation for the observed minus calculated (obs $-$ calc) wavenumbers of about 1 \cm, an order of 
magnitude improvement over the previous results.\cite{Koput2} An \ai\ line list with this accuracy  could be useful for a number of applications. 
However, for most applications it is also necessary to accurately compute  highly excited rotational levels.  This is done in this work. Before looking at high $J$ rotational levels, we reconsidered the $J=0$ results of  Ma\l yszek and  Koput \cite{Koput3} using both EKE programme WAVR4 \cite{jt339} and approximate kinetic energy programme TROVE \cite{trove-paper}.

Diatom-diatom HO--OH coordinates were employed in the programme WAVR4; these coordinates were one of those used to consider acetylene -- vinylidene isomerisation \cite{jt346}.  The calculations used a discrete variable
representations (DVR) based on a grid of 10 radial functions for each OH coordinate
and 18 radial functions for the OO coordinate. The parameters used for
OH stretch Morse-oscillator like functions were $r_{\rm e} = 0.91$ \AA,
$\omega_{\rm e} = 2500$ \cm\ and $D_{\rm e} = 35000$ \cm, and $r_{\rm e} =1.53$ \AA,
$\omega_{\rm e} = 1500$ \cm\ and $D_{\rm e} = 45 000$ \cm\ for the OO stretch.
The bending basis set consists of coupled angular functions
\cite{jt339} defined by $j^{\rm max}=l^{\rm max} =22$ and $k^{\rm max}$ =
12. The resulting energy levels with $J$ = 0 were within
0.1 \cm\ of the previous calculations \cite{Koput3}, see Table~\ref{tab:J1}.
However for WAVR4 calculations of the same accuracy for levels with  $J = 1$
require about 10 times more computer time. This is a consequence
of the $J$ -- $K$ coupling used in the EKE procedure. This coupling
is essential for the linear HCCH system \cite{jt346} and very floppy
molecules \cite{jt312}, but not for \ho.
The use of WAVR4 to calculate energies of high $J$ levels is
computationally unrealistic at present and we note that indeed    corresponding
studies on acetylene have thus far been confined to low $J$ values
\cite{jt479}.

TROVE is a computer suite for rovibrational calculations of energies and intensities for molecules of (at least in principle) arbitrary structures. TROVE uses a multilevel contraction scheme for constructing the rovibrational basis set. The primitive basis functions are given by products of six 1-dimensional (1D) functions $\phi_i(\xi_i)$, where $\xi_i$ represents one of the six internal coordinates. For HOOH we choose $\xi_1$, $\xi_2$,  and $\xi_3$ to be the linearized versions of the three stretching  internal displacements
$R-R_{\rm e}$, $r_1-r_{\rm e}$, and $r_2-r_{\rm e}$, respectively,
$\xi_{4}$ and $\xi_{5}$
are the linearized versions of the two bending
displacements $\theta_1-\theta_{\rm e}$
and $\theta_2-\theta_{\rm e}$; $\xi_6$ is
the torsional coordinate $\tau$, see Fig.~\ref{f:tau}.

The kinetic energy operator in TROVE is given by an
expansion in terms of the five coordinates $\xi_{i}$,
representing the rigid modes $i=1\ldots 5$. The
potential energy function is also expanded but using three Morse-type
expansion variables $1-\exp(-a_i\xi_i)$ ($i=1,2,3$) and two bending
coordinates $\xi_4$ and $\xi_5$.
Here $a_1 = 2.2$ \AA$^{-1}$, $a_2=a_3=2.3$ \AA$^{-1}$
were selected to match closely the shape of the \ai\
PES along the stretching modes. In the present work we
employ  6th and 8th order expansions to represent,
respectively, the  kinetic energy operator and potential
energy function.

The rovibrational motion of the non-rigid molecule HOOH is
best represented by the extended $C_{\rm 2h}^{+}$(M)
molecular symmetry group \cite{Hoxxxx.HOOH}, which is
isomorphic to $D_{\rm 2h}$(M) as well as to the extended
group $G(4)$(EM).\cite{BJ} As explained in detail by Bunker
and Jensen\cite{BJ}, the extended group is needed to describe
the torsional splitting due both the \textit{cis}-
and \textit{trans}-tunnelings. In the present work we use
the $D_{\rm 2h}$(M) group to classify the symmetry of the HOOH states.
This group is given by the eight irreducible
representations $A_{\rm g}, A_{\rm u}, B_{\rm 1g},
B_{\rm 1u}, B_{\rm 2g}, B_{\rm 2u}, B_{\rm 3g}, B_{\rm 3u}$.
In order to account for the extended symmetry properties of the
floppy HOOH molecules an extended range for torsion motion,
from 0$^{\circ}$ to 720$^\circ$
was introduced into TROVE. In this representation
$\tau = 0$ and 720$^\circ$ correspond to the \textit{cis} barrier, while at $\tau=360^{\circ}$ the molecule has the \textit{trans} configuration.


TROVE's primitive basis functions, $\phi_{v_i}^{(i)}$ ($i=1\ldots 6$),
are generated numerically by solving six 1D vibrational Schr\"{o}dinger
equations for each vibrational mode $i$ employing the
Numerov-Cooley method~\cite{Numerov,Cooley}. The corresponding
reduced 1D Hamiltonian operators $H^{\rm 1D}_i$
($i=1\ldots 6$) are obtained by freezing the five remaining modes at
the corresponding equilibriums. The integration ranges are
selected to be large enough to accommodate all basis functions
required (see below for the discussion of the basis set sizes).
The torsional functions are obtained initially on the
range $\tau=0\ldots 360^{\circ}$ and transform according
with the $C_{\rm 2h}$(M) group. A very fine grid of 30,000 points
and the quadruple numerical precision [real(16)] was used for
generating the eigenfunctions of the corresponding Schr\"{o}dinger
equation in order to resolve the $trans$ splittings up to $v_6=42$.
The wavefunctions are then extended to $\tau=360\ldots 720^\circ$
through the $+/-$ reflection of the $C_{\rm 2h}$(M) values and
classified according to $D_{\rm 2h}$(M). The extended primitive
torsion functions are able to account for the torsional splitting
due to both the \textit{cis} and \textit{trans} tunneling.

The 6D primitive basis functions are then formed from different
products of the 1D functions $\phi_{v_i}^{(i)}$. The size of the basis set
is controlled in TROVE by the so-called polyad number $P$, which in the present case is given by
\begin{equation}\label{e:polyad}
  P = 4 v_1 + 8 (v_2+v_3) + 8 (v_4 + v_5) + v_6 \le P_{\rm max},
\end{equation}
where $v_i$, $i=1\ldots 6$ are the local mode quantum numbers
corresponding to the primitive functions $\phi_{v_i}^{(i)}(\xi_{i})$.
The primitive basis set is then processed through a number of
contractions, as described in detail previously \cite{jt466},
to give a  final basis set in the $J=0$ representation.

Calculations using programme TROVE started with a search for a basis
set and operator expansions which would give results close to
the EKE ones.  The final values of the basis set parameters were
chosen to try to meet two conflicting requirements: the best possible
convergence and a compact enough calculation  to
allow high $J$ energy levels to be computed. The
final basis set parameters were the following: the maximum polyad
number $P_{\rm max}=42$, which also corresponds to the highest excitation of the torsional mode $v_6$.
For O-O stretch the maximal number was 8, for OH stretches - 8 and for
the bending modes - 10. These parameters control the size of the basis
set  used in the TROVE calculations according with Eq.~(\ref{e:polyad}). For the $J=0$ levels
they give good agreement with the previous studies, see Table~\ref{tab:J0}.

\section{Results}
\label{s:results}

The updated version of TROVE was used to calculate excited
rotational levels for $J$ up to 35, the highest assigned thus far experimentally.
Initial calculations were performed with the equilibrium distances and
angles obtained \ai\ in Ref. \citenum{Koput3}. In this case
the discrepancies between theory and experiment increased
quadratically with increasing $J$: the $J = 1$ levels
were calculated with an accuracy around 0.001 \cm, but those for $J = 35$
differ from experiment by about 1 \cm.

We therefore chose to adjust the equilibrium parameters $R_{\rm e}$
and $r_{\rm e}$ to better reproduce the experimental values.  Only
very small changes were needed to make the $J = 35$ levels accurate to
about 0.03 \cm\ for the ground vibrational state.  In particular, the
original value of $R_{\rm e}$ of 1.45539378 \AA\ was shifted to
1.45577728 \AA\ and $r_{\rm e} = 0.96252476$ \AA\ moved to 0.96253006 \AA.
The rotational structure
within the excited vibrational states is of similar accuracy, meaning
that these levels are essentially shifted just by the discrepancy in
the vibrational band origin.  In practice this geometry shift not only
meant that low $J$ energy levels were reproduced with an accuracy of 1
\cm, reproducing the accuracy of the vibrational band origin, but
also resulted in pseudo-resonance artifacts. To illustrate
this consider the interaction between the ground vibrational state and
the low-lying $v_4 = 1$ torsional vibrational state, which lies about
2 \cm\ too low in the calculations.  This results is an artificial
closeness, and interaction, between levels with the same $J$  and
$K_a= 8$ for $v=0$ and $K_a = 6$ for $v_4 = 1$.  The resulting shift
in the energy levels is significant; it grows with $J$ and reaches
about 1 \cm\ at $J$ = 30. We call this a  pseudo-resonance artifact
since no such interaction is seen in the experimentally-determined energy levels.

There are several ways to avoid this artificial pseudo-resonances.  One
would be to fit the PES to experimental data, which would remove this
artificial near-degeneracy. This is likely to be a topic of future
work. An alternative possibility, which is already available within
TROVE \cite{jt466},  is to simply adjust the calculated values of the
vibrational band origins given by the $J = 0$ calculation to the
observed ones prior to their use in calculations of the $J>0$ levels.
This option, which is not available in EKE codes which couple the
bending basis with the rotational functions \cite{jt14}, not only
shifts the energies, it also rearranges the matrix elements so that the
artificial resonances disappear.  With this adjusted calculation the
energy levels vary smoothly with the increasing $J$ and $K_a$ quantum
numbers, see Tables \ref{tab:J1}, \ref{tab:J30} and \ref{tab:J35}, as one would expect
\cite{jt205} from purely \ai\ levels.

One other problem remained when comparing our rotationally excited
energy levels with the observed ones.  This concerned rotational
levels with the quantum number $K_a = 1$ which did not behave as
levels associated with other values of $K_a$. The error for the
two levels with $K_a = 1$ increases disproportionately to that of
other levels as $J$ increases. This error was about 0.1 \cm\ for $J$ =
30. A series of test calculations revealed the reason for such
discrepant behaviour of the levels with $K_a = 1$. It transpires that a
small change in the height of the torsional barrier, which is strongly
influenced by the linear expansion coefficient $c_{000001}$, does not affect other $K_a$ levels, but significantly influences only those with $K_a$ = 1.
Varying this expansion coefficient can both increase and decrease the splitting of the $K_a$ = 1 doublet.  As this splitting is overestimated in calculations
using the \ai\ value of $c_{000001}$, its reduction by about 1~\%\ from
the \ai\ value of 0.00487 to 0.00483 $E_h$
results in roughly a fourfold   improvement of the obs -- calc value for the $K_a=1$ levels. This change affects the value of
the  ground-state torsional splitting of 11 \cm\ and also the values of the other torsional energy levels, all of which move significantly closer to the observed values, than the purely \ai\ levels given in the Table~\ref{tab:J0}.  In
particular, this small adjustment improves the calculated  ground-state splitting to 11.4 \cm\ and the first torsional level to 255.2 \cm.
Thus adjusting $c_{000001}$
not only improves  significantly the values of  levels with $K_a=1$ levels,
      it
improves the overall agreement with experiment for the band origins.               The
underlying reason for this is that the expansion coefficient  $c_{000001}$
controls the height of the torsional barrier.

These minor adjustments result in very accurate values for rotational
energy levels a sample of which are presented in Tables~\ref{tab:J1},
\ref{tab:J30} and \ref{tab:J35}. A more comprehensive set of energy
levels is given in the supplementary material \cite{supl}. From these
tables one can see that the discrepancy between observed and
calculated energy values increases both gently and smoothly with
rotational quantum number $J$. Such calculations therefore provide an
excellent starting point for assigning high $J$ transitions both
within the ground state and to excited vibrational states, as the
density of observed transitions is orders of magnitude smaller  than the accuracy of calculations.

\section{Conclusions}
\label{s:concl}

We present results of \ai\ and slightly adjusted \ai\ calculations for
the vibrational and rovibrational energy levels of the \ho\ molecule.
Use of the accurate \ai\ PES calculated by Ma\l yszek and Koput
\cite{Koput3} reproduces the known vibrational band origin with a
standard deviation of about 1 \cm.  The use of programme TROVE
\cite{Trove} for the nuclear motion calculations allowed us compute
high rotational levels up to $J=35$. Indeed,
energy levels with $J=50$
could be calculated on a high-end workstation, and the accuracy of prediction will be very
high - better than 0.5 \cm. However, we have not yet performed such calculations,
as no comparison with  experimental values is currently possible.
Experimentally derived energy levels up to $J=35$ are compared with our
calculations. These are reproduced with an unprecedented accuracy of
0.001 \cm\ for the levels up to $J=10$ and 0.02 \cm\ for all the known
levels above this.  Variational calculations using this
slightly adjusted \ai\ PES
results in very smooth variation in the discrepancies between the
observed and calculated levels as a function of the rotational quantum
numbers $J$ and $K_a$. This smoothness and accuracy is the key to the
successful analysis of previously unassignable spectra
\cite{jt200,jt205} as, in particular, the accidental resonances, which
seriously complicate any analysis based on the use an effective
Hamiltonian, are automatically allowed for in such calculations.

There is one other important aspect of the \ho\ rotation-vibration
problem which we should mention. The detection of extrasolar planets and,
in particular, our ability to probe  the molecular composition of these
bodies using spectroscopy \cite{jt400}, has led to demand for
accurate, comprehensive line lists over an extended range of both
temperature and wavelength for all species of possible importance in
exoplanet atmospheres \cite{jt528}. The accuracy
of the calculations  presented here and, especially their
ability to reliably predict highly excited rotational levels which
are of increasing importance at higher temperatures, suggests
that the present work will provide an excellent starting point
for the calculation of a comprehensive line list for \ho. In this
we will be following the recent work of Bowman and co-workers who have
computed similar line lists for somewhat more rigid hydrocarbon
systems \cite{09WaScSh.CH4,12CaShBo.C2H4}.

\section*{Acknowledgment}
This work was performed as part of ERC Advanced Investigator Project 267219.
We  also thank the Russian Fund for Fundamental Studies
for their support for aspects of this project.

\newpage

\begin{table}
\caption{ Calculated and observed energy levels, in \cm, for $J=0$  using WAVR4,  TROVE and
published by Ma{\l}yszek and Koput (MK) \cite{Koput3}; ``tr-shift'' results are computed with an adjusted height
for the torsional barrier. The observed values are taken from Ref.~\protect\onlinecite{perin2,perin3,perin4,95PeVaFl.H2O2}.}
\label{tab:J0}
\begin{tabular}{cccccccrrrrr}
\hline
\hline
$v_1$  & $v_2$ & $v_3$ &    $v_4$   & $v_5$    & $v_6$  & Sym&   Obs   &   WAVR4  &  TROVE &tr-shift & MK\\
\hline

0 & 0   &  0  &  1       &  0     &  0   & $A_{\rm g}$           &  254.550 &  256.406 & 256.419 & 255.490 & 255.43\\
 0 & 0   &  0  &  2       &  0     &  0  &  $A_{\rm g}$         &  569.743 &  570.334 & 570.251 & 570.690 & 570.45 \\
 0 & 0   &  1  &  0       &  0     &  0  &    $A_{\rm g}$      &  865.939 &  865.547 & 865.652 & 865.468  & 866.02\\
 0 & 0   &  0  &  3       &  0     &  0  &    $A_{\rm g}$      & 1000.882 & 1001.227 &1001.073 & 1002.493 & 1001.92 \\
 0 & 0   &  0  &  0       &  0     &  1  &  $B_{\rm u}$         & 1264.583 & 1264.819 &1265.121 & 1264.868  & 1264.54\\
 0 & 0   &  0  &  1       &  0     &  1  &   $B_{\rm u}$        & 1504.872 & 1505.977 &1506.283 & 1505.634 & \\
 0 & 0   &  0  &  2       &  0     &  1  &  $B_{\rm u}$                  & 1853.634 & 1853.949 &1854.424 & 1855.305 & \\
 0 & 0   &  0  &  0       &  0     &  0  &   $A_{\rm u}$       &   11.437 &   11.014 &  10.997 & 11.289 & 11.28 \\
 0 & 0   &  0  &  1       &  0     &  0  &     $A_{\rm u}$      &  370.893 &  371.247 & 371.203 & 371.478& 371.32 \\
 0 & 0   &  0  &  2       &  0     &  0  &    $A_{\rm u}$       &  776.122 &  776.465 & 776.320 & 777.336 & 776.93 \\
 0 & 0   &  1  &  0       &  0     &  0  &   $A_{\rm u}$        &  877.934 &  877.094 & 877.200 &  877.303 &  \\
 0 & 0   &  0  &  0       &  0     &  1  &  $B_{\rm g}$           & 1285.121 & 1284.889 &1285.249 &  1285.457&  \\
 0 & 0   &  0  &  1       &  0     &  1  &   $B_{\rm g}$          & 1648.367 & 1648.553 &1649.012 &  1649.485&  \\
 0 & 0   &  0  &  2       &  0     &  1  &   $B_{\rm g}$           & 2072.404 & 2072.384 &2072.949 &  2074.231&   \\

\hline
\hline
\end{tabular}

\end{table}

\begin{table}
\caption{Calculated and observed energy levels, in \cm, for
the vibrational ground state (left hand column) and the (000 100 $A_{\rm g}$) state - (right hand column)
 with $J$ = 1, 3 and 5. Observed energy levels taken from  Ref.~\protect\onlinecite{perin3}.}
\label{tab:J1}
\begin{tabular}{cccrrlrrl}
\hline
\hline
$J$  & $K_a$ & $K_c$ &    Obs   &    Calc    &  o-c     &   Obs     &   Calc   &  o-c   \\
1 & 0  & 1   &   1.71154 &    1.71152 &  0.00002 &   256.255 &  256.255 &  0.000 \\
1 & 1  & 1   &  10.90677 &   10.9068  &  0.0000  &   265.427 &  265.427 &  0.000 \\
1 & 1  & 0   &  10.9426  &   10.9426  &  0.0000  &   265.474 &  265.475 &  0.001 \\
  &    &     &           &            &          &           &          &        \\
3 & 0  & 3   &  10.2683  &   10.2682  &  0.0001  &   264.777 &  264.777 &  0.000 \\
3 & 1  & 3   &  19.374   &   19.374   &  0.000   &   273.830 &  273.831 & -0.001 \\
3 & 1  & 2   &  19.589   &   19.589   &  0.000   &   274.119 &  274.117 &  0.002 \\
3 & 2  & 2   &  47.115   &   47.115   &  0.000   &   301.555 &  301.556 &  0.001 \\
3 & 2  & 1   &  47.115   &   47.115   &  0.000   &   301.556 &  301.557 &  0.001 \\
3 & 3  & 1   &  93.155   &   93.155   &  0.000   &   347.509 &  347.512 &  0.003 \\
3 & 3  & 0   &  93.155   &   93.155   &  0.000   &   347.509 &  347.512 &  0.003 \\
  &    &     &           &            &          &           &          &        \\
5 & 0  & 5   &  25.6667  &   25.6664  &  0.0003  &   280.113 &  280.112 &  0.001 \\
5 & 1  & 5   &  34.613   &   34.613   &  0.000   &   288.954 &  288.954 &  0.000 \\
5 & 1  & 4   &  35.151   &   35.150   &  0.001   &   289.671 &  289.671 &  0.000 \\
5 & 2  & 4   &  62.513   &   62.513   &  0.000   &   316.893 &  316.893 &  0.000 \\
5 & 2  & 3   &  62.517   &   62.517   &  0.000   &   316.899 &  316.899 &  0.000 \\
5 & 3  & 3   & 108.551   &  108.551   &  0.000   &   362.845 &  362.847 &  0.002 \\
5 & 3  & 2   & 108.551   &  108.551   &  0.000   &   362.845 &  362.847 &  0.002 \\
5 & 4  & 2   & 172.968   &  172.968   &  0.000   &   427.143 &  427.146 &  0.003 \\
5 & 4  & 1   & 172.968   &  172.968   &  0.000   &   427.143 &  427.146 &  0.003 \\
5 & 5  & 1   & 255.733   &  255.733   &  0.000   &   509.757 &  509.764 &  0.007 \\
5 & 5  & 0   & 255.733   &  255.733   &  0.000   &   509.757 &  509.764 &  0.007 \\
\hline
\hline
\end{tabular}

\end{table}

\begin{table}
\caption{Calculated and observed energy levels, in \cm, for
the (000 200    $A_{\rm g}$) (left hand column) and the (000 300   $A_{\rm g}$) state - right hand column)
 with $J$ = 1, 3 and 5. Observed energy levels taken from Ref.~\protect\onlinecite{perin3}.}
\begin{tabular}{cccrrlrrl}
\hline
\hline
$J$  & $K_a$ & $K_c$ &    Obs   &    Calc    &  o-c     &   Obs     &   Calc   &  o-c   \\
1 & 0  & 1   & 571.448   &  571.449   & -0.001   &  1002.584 & 1002.584 &  0.000 \\
1 & 1  & 1   & 580.550   &  580.549   &  0.001   &  1011.664 & 1011.614 &  0.050 \\
1 & 1  & 0   & 580.577   &  580.576   &  0.001   &  1011.678 & 1011.628 &  0.050 \\
  &    &     &           &            &          &           &          &        \\
3 & 0  & 3   & 579.976   &  579.975   &  0.001   &  1011.098 & 1011.097 &  0.001 \\
3 & 1  & 3   & 589.009   &  589.007   &  0.002   &  1020.143 & 1020.093 &  0.050 \\
3 & 1  & 2   & 589.174   &  589.172   &  0.002   &  1020.225 & 1020.176 &  0.049 \\
3 & 2  & 2   & 616.427   &  616.427   &  0.000   &  1047.246 & 1047.244 &  0.002 \\
3 & 2  & 1   & 616.427   &  616.427   &  0.000   &  1047.246 & 1047.244 &  0.002 \\
3 & 3  & 1   & 661.974   &  661.975   &  0.001   &  1092.462 & 1092.411 &  0.051 \\
3 & 3  & 0   & 661.974   &  661.975   &  0.001   &  1092.462 & 1092.411 &  0.051 \\
  &    &     &           &            &          &           &          &        \\
5 & 0  & 5   & 595.324   &  595.324   &  0.000   &  1026.421 & 1026.021 &  0.008 \\
5 & 1  & 5   & 604.235   &  604.232   &  0.003   &  1035.404 & 1035.354 &  0.050 \\
5 & 1  & 4   & 604.645   &  604.644   &  0.001   &  1035.610 & 1035.561 &  0.049 \\
5 & 2  & 4   & 631.773   &  631.773   &  0.000   &  1062.566 & 1062.565 &  0.001 \\
5 & 2  & 3   & 631.775   &  631.775   &  0.000   &  1062.566 & 1062.565 &  0.001 \\
5 & 3  & 3   & 677.317   &  677.315   &  0.002   &  1107.779 & 1107.729 &  0.050 \\
5 & 3  & 2   & 677.317   &  677.315   &  0.002   &  1107.779 & 1107.729 &  0.050 \\
5 & 4  & 2   & 741.046   &  741.044   &  0.002   &  1170.935 & 1170.930 &  0.005 \\
5 & 4  & 1   & 741.046   &  741.044   &  0.002   &  1170.935 & 1170.930 &  0.005 \\
5 & 5  & 1   & 822.934   &  822.930   &  0.004   &  1252.191 & 1252.140 &  0.051 \\
5 & 5  & 0   & 822.934   &  822.930   &  0.004   &  1252.191 & 1252.140 &  0.051 \\
\hline
\hline
\end{tabular}

\end{table}

\begin{table}
\caption{Calculated and observed energy levels, in \cm, for
the (001 000    $A_{\rm g}$) (left hand column) and the (000 000 $A_{\rm u}$)
 state - right hand column)
 with $J$ = 1, 3 and 5. Observed energy levels taken from  Ref.~\protect\onlinecite{perin3}.}
\begin{tabular}{cccrrlrrl}
\hline
\hline
$J$  & $K_a$ & $K_c$ &    Obs   &    Calc    &  o-c     &   Obs     &   Calc   &  o-c   \\
1 & 0  &  1  &  867.628  &  867.628   &  0.000   &    13.150 &   13.149 &  0.001 \\
1 & 1  &  0  &  876.815  &  876.816   &  0.001   &    22.337 &   22.337 &  0.000 \\
1 & 1  &  0  &  876.851  &  876.850   &  0.001   &    22.369 &   21.368 &  0.001 \\
  &    &     &           &            &          &           &          &        \\
3 & 0  &  3  &  876.074  &  875.073   &  0.001   &    21.712 &   21.711 &  0.001 \\
3 & 1  &  3  &  885.171  &  884.171   &  0.000   &    30.821 &   30.820 &  0.001 \\
3 & 1  &  2  &  885.386  &  885.387   & -0.001   &    31.009 &   31.009 &  0.000 \\
3 & 2  &  2  &  912.886  &  912.885   &  0.001   &    58.518 &   58.518 &  0.000 \\
3 & 2  &  1  &  912.887  &  912.887   &  0.000   &    58.518 &   58.518 &  0.000 \\
3 & 3  &  1  &  958.884  &  958.887   & -0.003   &   104.508 &  104.508 &  0.000 \\
3 & 3  &  0  &  958.884  &  958.887   & -0.003   &   104.508 &  104.508 &  0.000 \\
  &    &     &           &            &          &           &          &        \\
5 & 0  &  5  &  891.273  &  891.2712  &  0.002   &    37.121 &   37.120 &  0.001 \\
5 & 1  &  5  &  900.210  &  900.209   &  0.001   &    46.089 &   46.088 &  0.001 \\
5 & 1  &  4  &  900.749  &  900.748   &  0.001   &    46.561 &   46.560 &  0.001 \\
5 & 2  &  4  &  928.085  &  928.085   &  0.000   &    73.926 &   73.926 &  0.000 \\
5 & 2  &  3  &  928.089  &  928.088   &  0.001   &    73.929 &   73.928 &  0.001 \\
5 & 3  &  3  &  974.080  &  974.082   & -0.002   &   119.913 &  119.913 &  0.000 \\
5 & 3  &  2  &  974.080  &  974.082   & -0.002   &   119.913 &  119.913 &  0.000 \\
5 & 4  &  2  & 1038.438  & 1038.440   & -0.002   &   184.260 &  184.261 & -0.001 \\
5 & 5  &  1  & 1121.127  & 1121.133   & -0.006   &   266.936 &  266.938 & -0.002 \\
\hline
\hline
\end{tabular}

\end{table}

\begin{table}
\caption{Calculated and observed energy levels, in \cm, for
the (000 100    $A_{\rm u}$) (left hand column) and the (000 200    $A_{\rm u}$)
state - (right hand column)
 with $J$ = 1, 3 and 5. Observed energy levels taken from the Ref.~\protect\onlinecite{perin3}.}
\begin{tabular}{cccrrlrrl}
\hline
\hline
$J$  & $K_a$ & $K_c$ &    Obs   &    Calc    &  o-c     &   Obs     &   Calc   &  o-c   \\
1 & 0  &  0  &  372.601  &  372.602   &  0.001   &   777.826 &  777.827 & -0.001 \\
1 & 1  &  1  &  381.733  &  381.733   &  0.000   &   786.881 &  786.891 & -0.010 \\
1 & 1  &  1  &  381.764  &  381.764   &  0.000   &   786.901 &  786.911 & -0.010 \\
  &    &     &           &            &          &           &          &        \\
3 & 0  &  3  &  381.140  &  381.139   &  0.001   &   786.350 &  786.350 &  0.000 \\
3 & 1  &  3  &  390.195  &  390.195   &  0.000   &   795.355 &  795.365 & -0.010 \\
3 & 1  &  2  &  390.378  &  390.376   &  0.002   &   795.472 &  795.482 & -0.010 \\
3 & 2  &  2  &  417.723  &  417.722   &  0.001   &   822.639 &  822.640 & -0.001 \\
3 & 2  &  1  &  417.723  &  417.722   &  0.001   &   822.639 &  822.640 & -0.001 \\
3 & 3  &  1  &  463.434  &  463.434   &  0.000   &   867.976 &  867.986 & -0.010 \\
3 & 3  &  0  &  463.434  &  463.434   &  0.000   &   867.976 &  867.986 & -0.010 \\
  &    &     &           &            &          &           &          &        \\
5 & 0  &  5  &  396.505  &  396.504   &  0.001   &   801.689 &  801.689 &  0.000 \\
5 & 1  &  5  &  405.426  &  405.423   &  0.003   &   810.607 &  810.617 & -0.010 \\
5 & 1  &  4  &  405.881  &  405.879   &  0.002   &   810.900 &  810.909 & -0.009 \\
5 & 2  &  4  &  433.088  &  433.088   &  0.000   &   837.977 &  837.977 &  0.000 \\
5 & 2  &  3  &  433.090  &  433.089   &  0.001   &   837.977 &  837.977 &  0.000 \\
5 & 3  &  3  &  478.796  &  478.794   &  0.002   &   883.310 &  883.320 & -0.010 \\
5 & 3  &  2  &  478.796  &  478.794   &  0.002   &   883.310 &  883.320 & -0.010 \\
5 & 4  &  2  &  542.756  &  542.756   &  0.000   &   946.769 &  946.770 &  0.001 \\
5 & 5  &  1  &  624.938  &  624.937   &  0.001   &  1028.289 & 1028.299 & -0.010 \\
\hline
\hline
\end{tabular}
\end{table}

\begin{table}
\caption{Calculated and observed energy levels, in \cm, for $J=30$. Results are for
the ground vibrational state (left hand column) and the (000 100    $A_{\rm g}$)
state - (right hand column). Observed energy levels taken from the Ref.~\protect\onlinecite{perin3}.}
\label{tab:J30}
\begin{tabular}{cccrrlrrl}
\hline
\hline
$J$  & $K_a$ & $K_c$ &    Obs   &    Calc    &  o-c     &   Obs     &   Calc   &  o-c   \\
\hline
30 & 0   & 30  &  789.577 &  789.581 & -0.004 & 1038.983 & 1038.925 &  0.058 \\
30 & 1   & 30  &  793.053 &  793.065 & -0.012 & 1041.540 & 1041.481 &  0.039 \\
30 & 1   & 29  &  809.594 &  809.565 &  0.029 & 1063.053 & 1063.007 &  0.046 \\
30 & 2   & 29  &  829.292 &  829.282 &  0.010 & 1080.396 & 1080.345 &  0.051 \\
30 & 2   & 28  &  832.547 &  832.521 &  0.026 & 1086.079 & 1086.036 &  0.043 \\
30 & 3   & 28  &  876.030 &  876.017 &  0.013 & 1127.862 & 1127.815 &  0.047 \\
30 & 3   & 27  &  876.191 &  876.174 &  0.017 & 1128.304 & 1128.262 &  0.042 \\
30 & 4   & 27  &  940.027 &  940.015 &  0.012 & 1191.838 & 1191.794 &  0.044 \\
30 & 4   & 26  &  940.029 &  940.013 &  0.016 & 1191.850 & 1191.807 &  0.043 \\
30 & 5   & 26  & 1022.323 & 1022.311 &  0.012 & 1274.332 & 1274.293 &  0.039 \\
30 & 5   & 25  & 1022.324 & 1022.312 &  0.012 & 1274.332 & 1274.293 &  0.039 \\
30 & 6   & 25  & 1122.848 & 1122.837 &  0.011 & 1371.047 & 1371.024 &  0.023 \\
30 & 6   & 24  & 1122.849 & 1122.838 &  0.011 & 1371.047 & 1371.024 &  0.023 \\
30 & 7   & 24  & 1241.304 & 1241.294 &  0.010 & 1491.957 & 1491.933 &  0.024 \\
30 & 7   & 23  & 1241.304 & 1241.294 &  0.010 & 1491.957 & 1491.933 &  0.024 \\
30 & 8   & 23  & 1381.938 & 1381.916 &  0.022 & 1628.749 & 1628.731 &  0.018 \\
30 & 9   & 21  & 1534.588 & 1534.578 &  0.010 & 1782.735 & 1782.739 & -0.004 \\
30 & 10  & 21  & 1707.358 & 1707.349 &  0.009 & 1958.577 & 1958.558 &  0.019 \\
30 & 11  & 19  & 1898.169 & 1898.159 &  0.010 & 2148.182 & 2148.186 & -0.004 \\
\hline
\hline
\end{tabular}
\end{table}

\begin{table}
\caption{Calculated and observed energy levels, in \cm, for $J=30$ levels of the (000 000    $A_{\rm u}$)
vibrational state. Observed energy levels taken from Ref.~\protect\onlinecite{perin3}.}
\begin{tabular}{cccrrl}
\hline
\hline
$J$  & $K_a$ & $K_c$ &    Obs   &    Calc    &  o-c     \\
\hline
30 & 0   & 30  &  802.349 &  802.340 &  0.009         \\
30 & 1   & 30  &  806.280 &  806.276 &  0.004         \\
30 & 1   & 29  &  820.768 &  820.746 &  0.022         \\
30 & 2   & 29  &  841.344 &  841.330 &  0.013         \\
30 & 2   & 28  &  843.911 &  843.892 &  0.019         \\
30 & 3   & 28  &  887.854 &  887.839 &  0.015         \\
30 & 3   & 27  &  887.964 &  887.949 &  0.015         \\
30 & 4   & 27  &  951.826 &  951.815 &  0.011         \\
30 & 4   & 26  &  951.828 &  951.817 &  0.011         \\
30 & 5   & 26  & 1034.096 & 1034.085 &  0.011         \\
30 & 5   & 25  & 1034.096 & 1034.085 &  0.011         \\
30 & 5   & 25  & 1134.609 & 1134.599 &  0.010         \\
30 & 5   & 24  & 1253.291 & 1253.284 &  0.007         \\
30 & 5   & 24  & 1390.060 & 1390.055 &  0.005         \\
30 & 5   & 23  & 1544.803 & 1544.803 &  0.000         \\
30 & 5   & 23  & 1717.164 & 1717.167 & -0.003         \\
\hline
\hline
\end{tabular}
\end{table}

\begin{table}
\caption{   Variationally calculated and  observed or predicted
using effective Hamiltonian)  energy levels, in \cm, for $J=35$. Observed energy levels taken from the Ref.~\protect\onlinecite{perin3}.}
\label{tab:J35}
\begin{tabular}{cccrrlrrl}
\hline
\hline
$J$  & $K_a$ & $K_c$ &    Obs   &    Calc    &  o-c     &   Obs     &   Calc   &  o-c   \\
\hline
35 & 0   & 35  & 1067.027 & 1067.037 & -0.010   & 1314.205 & 1314.124 &  0.081  \\
35 & 1   & 35  & 1069.466 & 1069.484 & -0.018   & 1315.869 & 1315.787 &  0.082  \\
35 & 1   & 34  & 1091.715 & 1091.680 &  0.035   & 1344.395 & 1344.338 &  0.057  \\
35 & 2   & 34  & 1108.893 & 1108.882 &  0.011   & 1358.709 & 1358.637 &  0.052  \\
35 & 2   & 33  & 1114.564 & 1114.523 &  0.041   & 1368.256 & 1368.203 &  0.053  \\
35 & 3   & 33  & 1156.257 & 1156.241 &  0.016   & 1407.374 & 1407.309 &  0.065  \\
35 & 3   & 32  & 1156.655 & 1156.628 &  0.027   & 1408.434 & 1408.382 &  0.052  \\
35 & 4   & 32  & 1220.062 & 1220.044 &  0.018   & 1471.314 & 1471.252 &  0.062  \\
35 & 4   & 31  & 1220.070 & 1220.051 &  0.019   & 1471.355 & 1471.296 &  0.059  \\
35 & 5   & 31  & 1302.102 & 1302.088 &  0.014   & 1553.875 & 1553.821 &  0.054  \\
35 & 5   & 30  & 1302.114 & 1302.098 &  0.016   & 1553.876 & 1555.822 &  0.054  \\
35 & 6   & 30  & 1402.366 & 1402.353 &  0.013   & 1648.318 & 1648.288 &  0.030  \\
35 & 6   & 29  & 1402.369 & 1402.350 &  0.019   & 1648.318 & 1648.288 &  0.030  \\
35 & 7   & 29  & 1520.294 & 1520.282 &  0.012   & 1769.879 & 1769.844 &  0.035  \\
35 & 8   & 28  & 1662.739 & 1662.707 &  0.032   & 1906.483 & 1906.459 &  0.024  \\
35 & 9   & 27  & 1814.222 & 1814.207 &  0.015   &          &          &         \\
35 & 10  & 26  & 1986.565 & 1986.552 &  0.013   &          &          &         \\
\hline
\hline
\end{tabular}
\end{table}

\begin{table}
\caption{   Variationally calculated and  observed or predicted using effective Hamiltonian)  energy levels, in \cm, for $J=35$}
\begin{tabular}{cccrrlrrl}
\hline
\hline
$J$  & $K_a$ & $K_c$ &    Obs   &    Calc    &  o-c     &   Obs     &   Calc   &  o-c   \\
\hline
35 & 0   & 35  & 1933.163 & 1932.118 &  0.045   & 1080.587 & 1080.574 &  0.013  \\
35 & 1   & 35  & 1936.042 & 1935.996 &  0.046   & 1083.472 & 1083.465 &  0.007  \\
35 & 1   & 34  & 1955.496 & 1955.415 &  0.081 u & 1102.961 & 1102.929 &  0.032  \\
35 & 2   & 34  & 1973.810 & 1973.743 &  0.067 u & 1121.270 & 1121.247 &  0.023  \\
35 & 2   & 33  & 1978.296 & 1978.228 &  0.068 u & 1125.774 & 1125.743 &  0.031  \\
35 & 3   & 33  & 2020.807 & 2020.749 &  0.058 u & 1168.268 & 1168.245 &  0.023  \\
35 & 3   & 32  & 2021.078 & 2021.020 &  0.058 u & 1168.540 & 1168.517 &  0.023  \\
35 & 4   & 32  & 2084.733 & 2084.691 &  0.042 u & 1232.076 & 1232.060 &  0.016  \\
35 & 4   & 31  & 2084.738 & 2084.768 &  0.042 u & 1232.081 & 1232.055 &  0.026  \\
35 & 5   & 31  & 2165.503 & 2165.420 &  0.083 u & 1314.156 & 1314.138 &  0.018  \\
35 & 5   & 30  & 2165.503 & 2165.420 &  0.083 u & 1314.156 & 1314.138 &  0.018  \\
   &     &     &          &          &          & 1414.475 & 1414.459 &  0.016  \\
   &     &     &          &          &          & 1532.940 & 1532.928 &  0.012  \\
   &     &     &          &          &          & 1669.447 & 1669.438 &  0.009  \\
   &     &     &          &          &          & 1823.841 & 1823.839 &  0.002  \\
\hline
\hline
\end{tabular}
\end{table}

\bibliographystyle{rsc}

\providecommand*{\mcitethebibliography}{\thebibliography}
\csname @ifundefined\endcsname{endmcitethebibliography}
{\let\endmcitethebibliography\endthebibliography}{}

\newpage

\end{document}